# Interplay of electron-phonon coupling, pseudogap, and superconductivity in CsCa$_2$Fe$_4$As$_4$F$_2$ studied using ultrafast optical spectroscopy

Qi-Yi Wu,[1, 2] Chen Zhang,[2] Bai-Zhuo Li,[3, 4] Hao Liu,[2] Jiao-Jiao Song,[2] Bo Chen,[2] Hai-Yun Liu,[5] Yu-Xia Duan,[2] Jun He,[2] Jun Liu,[1] Guang-Han Cao,[3] and Jian-Qiao Meng[2, *]

[1] *School of Materials Science and Engineering, Central South University, Changsha 410083, Hunan, China*
[2] *School of Physics, Central South University, Changsha 410083, Hunan, China*
[3] *School of Physics, Zhejiang University, Hangzhou 310058, Zhejiang, China*
[4] *School of Physics and Optoelectronic Engineering, Shandong University of Technology, Zibo 255000, P. R. China*
[5] *Beijing Academy of Quantum Information Sciences, Beijing 100085, China*
(Dated: Wednesday 12$^{\text{th}}$ March, 2025)

The quasiparticle relaxation dynamics of the iron-based superconductor CsCa$_2$Fe$_4$As$_4$F$_2$ ($T_c \sim 29$ K) were investigated using ultrafast optical spectroscopy. A pseudogap ($\Delta_{PG} = 3.3 \pm 0.3$ meV) was observed on set below $T^* \approx 60$ K, prior to the emergence of a superconducting gap ($\Delta(0) = 6.6 \pm 0.4$ meV). At high excitation fluence, a coherent $A_{1g}$ phonon mode at 5.49 THz was identified, exhibiting deviations from anharmonic behavior below $T_c$. The electron-phonon coupling constant for this mode was estimated to be $\lambda_{A_{1g}} \approx 0.23$. These results provide insights into the interplay between the electron-phonon interactions, pseudogap, and the superconducting pairing mechanism in CsCa$_2$Fe$_4$As$_4$F$_2$.

Unconventional superconductivity, particularly in high-temperature superconductors (HTSCs), presents a significant challenge in condensed matter physics, as the underlying microscopic mechanisms remain elusive. The complex interplay of charge, orbital, and spin degrees of freedom in these strongly correlated systems complicates the identification of the critical factors governing superconductivity [1]. A prominent focus of ongoing research is the pseudogap phase observed in the normal state of numerous unconventional superconductors. While the gap opening temperature and the superconducting critical temperature ($T_c$) exhibit a clear correspondence in conventional superconductors, unconventional superconductors such as cuprates display a gap opening at a significantly higher temperature than $T_c$ [2]. However, the existence of a pseudogap in iron-based superconductors (FeSCs) remains controversial [3–9], with experimental evidence supporting both its presence and absence. It is imperative to resolve this ambiguity, as the pseudogap may be intimately linked to the pairing mechanism. Additionally, the nature of the electron-boson coupling responsible for pairing in unconventional superconductors is a subject of intense debate. In contrast to conventional superconductors, where electron-phonon coupling provides the pairing "glue," unconventional superconductors may rely on alternative mechanisms, such as antiferromagnetic magnons, spin fluctuations, or orbital fluctuations, as well as phonons. Although electron-phonon (e-ph) coupling has historically been considered too weak to account for high-temperature superconductivity [10–12], recent studies indicate that it may play a more substantial role than previously thought [13–16]. Addressing these fundamental issues is critical for unlocking the mechanisms underlying unconventional superconductivity.

Recently, the family of stoichiometric quasi-two-dimensional FeSCs, $A$Ca$_2$Fe$_4$As$_4$F$_2$ ($A$ = K, Cs, and Rb), have garnered significant attention due to its high $T_c$ ($\sim 30$ K), structural similarities to cuprates, and the absence of both magnetic and structural orders [17–19]. These compounds feature a complex structure, depicted in Fig. 1(a), that can be viewed as an intergrowth of 122-type $A$Fe$_2$As$_2$ and 1111-type CaFeAsF. This results in double FeAs layers sandwiched between insulating Ca$_2$F$_2$ layers, reminiscent of the bilayer structure in cuprates. The nature of the superconducting gap symmetry in these materials remains a subject of debate. While specific heat measurements [20, 21], optical spectroscopy [22], and angle-resolved photoemission spectroscopy (ARPES) experiments [23] suggest a nodeless gap structure, muon spin rotation experiments indicate the presence of line nodes in the gap [24, 25]. Evidence for a pseudogap in $A$Ca$_2$Fe$_4$As$_4$F$_2$ compounds has been reported through various experimental techniques, including NMR [26], optical conductivity [27], Nernst effect [28], transport measurements [19, 29], and scanning tunneling microscopy (STM) [30], although it is not confirmed by ARPES results [23, 31].

In this Letter, we employ ultrafast optical spectroscopy to investigate CsCa$_2$Fe$_4$As$_4$F$_2$ ($T_c \sim 29$ K). The transient reflectivity ($\Delta R/R$) exhibits distinct behavior above and below $T_c$. Above $T_c$, $\Delta R/R$ is well described by a bi-exponential decay, while an additional slow decay component emerges below $T_c$, indicating hindered quasiparticle relaxation due to the superconducting gap. Moreover, the faster decay process exhibits anomalies below $T^* \approx 60$ K, suggesting the possible emergence of a pseudogap. Under high excitation fluence, coherent oscillations associated with the $A_{1g}$ phonon mode (5.49 THz) were observed. The temperature dependence of this mode's frequency deviates from the expected anharmonic behavior, showing a downturn below $T_c$. The estimated nominal $e$-ph coupling constant for this mode is $\lambda_{A_{1g}} \approx 0.23$. These findings suggest a potential connection between the $A_{1g}$ phonon mode, the pseudogap, and the superconducting pairing mechanism in CsCa$_2$Fe$_4$As$_4$F$_2$.

High-quality single crystals of CsCa$_2$Fe$_4$As$_4$F$_2$ with well-defined (001) cleavage planes were grown using the self-flux method [19]. Ultrafast optical spectroscopy measurements were performed using 800 nm ($\sim 1.55$ eV) laser pulses with a 35 fs pulse width and 1 MHz repetition rate [32–34].

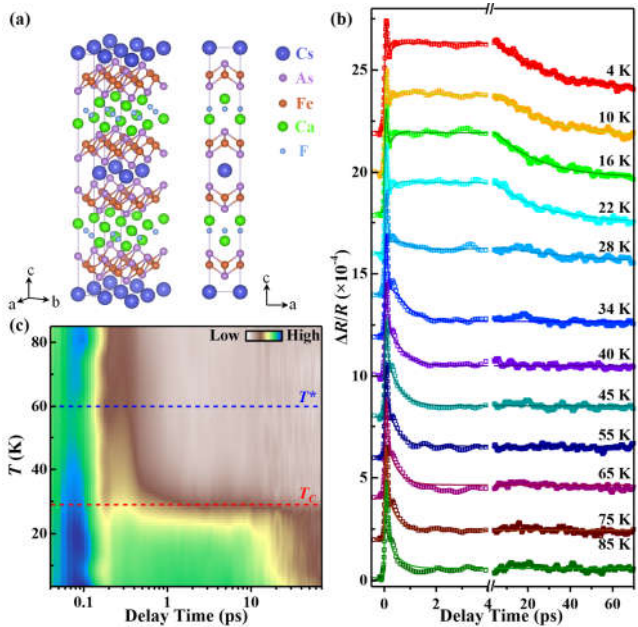

FIG. 1. (a) Crystal structure of $CsCa_2Fe_4As_4F_2$. (b) Transient reflectivity ($\Delta R/R$) as a function of delay time at various temperatures, measured at a pump fluence of $\sim 4.97\ \mu J/cm^2$. The solid lines are Eq. (1) fits. Note the break in the $x$ axis. (c) 2D pseudocolor map of $\Delta R/R$ as a function of temperature and delay time. The dashed red and blue lines indicate $T_c$ and $T^*$, respectively.

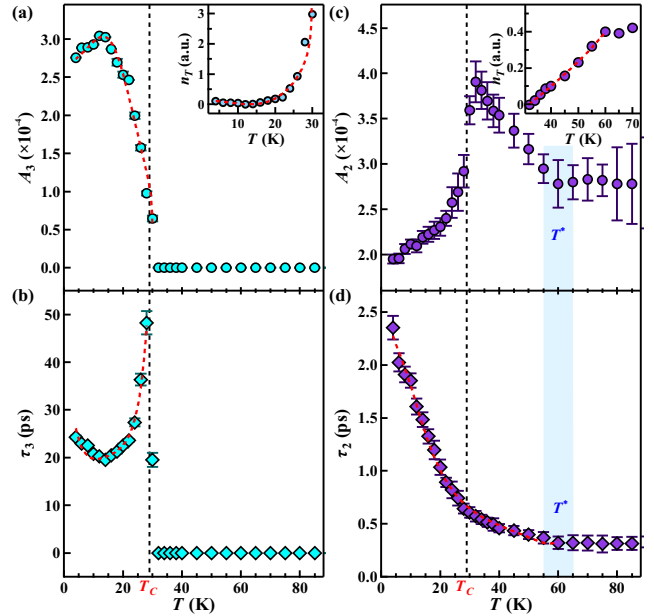

FIG. 2. Temperature dependence of the amplitudes ($A_i$) and relaxation times ($\tau_i$) for the(a), (b) third and (c),(d) second decay processes. The insets in (a) and (c) show the density of thermally excited quasiparticles ($n_T$). Fits to the data using the RT model are indicated by the red dashed lines, as detailed in the main text.

The pump and probe pulses were orthogonally polarized to enhance the signal-to-noise ratio of the transient reflectivity ($\Delta R/R$) measurements. Experiments were conducted on freshly cleaved sample surfaces under high vacuum ($10^{-6}$ mbar) over a temperature range of 4 K to 300 K.

Figure 1(b) presents the typical reflectivity signal, $\Delta R/R$, as function of delay time for various temperatures between 4 and 85 K. These measurements were conducted at a low pump fluence ($\sim 4.97\ \mu J/cm^2$) to avoid perturbing the superconducting state. Upon arrival of the pump pulse, $\Delta R/R$ exhibits a sharp rise, followed by a picosecond-timescale relaxation towards equilibrium. In particular, after the initial relaxation, $\Delta R/R$ shows a pronounced increase below $T_c$. Figure 1(c) displays a two-dimensional (2D) pseudocolor map of $\Delta R/R$ as a function of delay time and temperature. The delay time axis is plotted logarithmically to emphasize the rapid initial dynamics. Notably, noticeable changes are also observed at a higher temperature around 60 K.

The transient reflectivity is primarily governed by electron-electron ($e$-$e$) and electron-boson scattering processes. To analyze the relaxation dynamics in $CsCa_2Fe_4As_4F_2$, the $\Delta R/R$ data were fitted using an exponential decay model convoluted with a Gaussian laser pulse (see Fig. S1 in the supplementary material [35] for details):

$$\frac{\Delta R(t)}{R} = \frac{1}{\sqrt{2\pi}w}\exp(-\frac{t^2}{2w^2}) \otimes [\sum_{i=1}^{3} A_i\exp(-\frac{t-t_0}{\tau_i})] + C,$$

$$(1)$$

where $A_i$ and $\tau_i$ represent the amplitude and relaxation time of the $i$th decay process, respectively. $w$ is the incidence pulse temporal duration, and $C$ is a constant offset representing a long-lived decay process. The initial and fastest relaxation process ($i = 1$), with a lifetime $\tau_1$ of a timescale of tens of femtoseconds, is generally attributed to $e$-$e$ scattering. This assignment is supported by the linear dependence of its amplitude ($\Delta R_{max}/R$) on pump fluence at various temperatures [36] (for details, see the Supplemental Material [35]).

Figure 2 summarizes the temperature dependence of the amplitudes ($A_i$) and relaxation times ($\tau_i$) extracted from the fits of Eq. (1) to the $\Delta R/R$ data. The third decay process is associated with superconducting transitions, with a relaxation time on the order of tens of picoseconds, occurring within the superconducting states. As shown in Fig. 2(a), its amplitude $A_3$ increases sharply below $T_c$. The relaxation time $\tau_3$ shows a divergence near $T_c$ and a slight upturn at lower temperatures [Fig. 2(b)]. These features can be quantitatively analyzed using the Rothwarf-Taylor (RT) model [37], which describes the relaxation dynamics of photoexcited quasiparticles in systems with a narrow energy gap near the Fermi level. Photoexcitation of a superconductor with a gap $\Delta$ generates high-frequency bosons ($> 2\Delta$) during quasiparticle relaxation. These bosons can re-excite quasiparticles, creating a bottleneck effect that slows down relaxation until the bosons decay or diffuse. Under the weak perturbation and strong bottleneck conditions, the RT model allows for the extraction of the gap size using the following relations [38–40]:

$$A(T) \propto \frac{\varepsilon_I[\Delta + k_B T/2]}{1 + \gamma\sqrt{2k_B T/\pi\Delta}e^{-\Delta/k_B T}},$$

$$(2)$$

$$\tau^{-1}(T) \propto [\frac{\delta}{\beta n_T + 1} + 2n_T](\Delta + \alpha T\Delta^4)$$

$$(3)$$

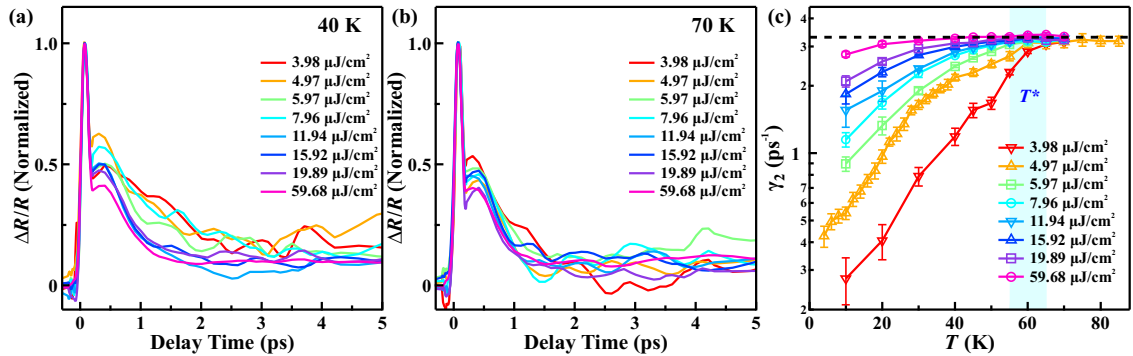

FIG. 3. (a) and (b) Normalized $\Delta R/R$ at 40 and 70 K, respectively, as a function of pump fluence. (c) Temperature dependence of the relaxation rate $\gamma_2$ ($1/\tau$) for different pump fluences, highlighting an anomaly around $T^*$.

$$n_T(T) = \frac{A(0)}{A(T)} - 1 \propto (T\Delta)^p e^{-\Delta/T} \quad (4)$$

where $\varepsilon_I$ is the absorbed laser energy density per unit cell, $n_T$ is the thermal quasiparticle density, and $\alpha$, $\beta$, $\gamma$, and $\delta$ are fitting parameters. The parameter $p$ ($0 < p < 1$) reflects the gapped density of states (DOS). This model has been successfully employed to explain the behavior of correlated systems, such as HTSCs [32, 33, 36–39] and heavy fermion systems [40–43]. Assuming a BCS-like temperature dependence for the superconducting gap [38, 39], a good fit to $A_3(T)$ [Fig. 2(a)], $n_T(T)$ [inset of Fig. 2(a)], $\tau_3(T)$ [Fig. 2(b)] is achieved, yielding a zero-temperature superconducting gap of $\Delta(0) = 6.6 \pm 0.4$ meV with $p = 0.5$. This value is consistent with the larger superconducting gaps observed in ARPES [31] and optical spectroscopy [22], while dirty-limit conditions have been suggested [22]. The corresponding ratio $2\Delta/k_B T_c \simeq 5.28$ significantly exceeds the weak-coupling BCS value of 3.52, indicating the presence of strong-coupling Cooper pairs.

In addition to the pronounced impact of the superconducting transition, the quasiparticle (QP) relaxation dynamics exhibit a subtle slowdown on a timescale of 0.3 - 2 ps below $T^* \approx 60$ K, as evident in Fig. 1(c). Figures 2(c) and 2(d) show the temperature dependence of the amplitude ($A_2$) and relaxation time ($\tau_2$) for the second decay process. Both $A_2$ and $\tau_2$ display anomalies around $T^*$, suggesting a change in the nature of the normal state at this temperature. Above $T^*$, $\tau_2$ remains relatively constant, while below $T^*$, it increases steadily with decreasing temperature. This trend is further supported by the temperature dependence data at various pump fluences in Fig. 3(c). $A_2$, on the other hand, initially increases between $T^*$ and $T_c$, but then decreases significantly in the superconducting state. This behavior indicating a suppression of the normal-state signal below $T_c$.

To further investigate the behavior around $T^*$, the fluence dependence of the transient reflectivity was examined. Figures 3(a) and 3(b) show the normalized $\Delta R/R$ at various pump fluences for temperatures of 40 K (below $T^*$) and 70 K (above $T^*$), respectively. At 40 K, the normalized curves exhibit a clear fluence dependence, becoming steeper with increasing fluence on the timescale of 0.3 to 2 ps and converging at high fluences. This fluence dependence vanishes at 70 K. Figure 3(c) displays the temperature dependence of the relaxation rate $\gamma_2$ (= $1/\tau_2$) for different pump fluences, revealing a significant fluence dependence below $\sim 60$ K that diminishes at higher temperatures.

This fluence-dependent behavior is a hallmark of systems with a pseudogap [44, 45], and its presence in $CsCa_2Fe_4As_4F_2$ provides strong evidence for the existence of a pseudogap phase below $T^*$. This behavior can be understood within the framework of the RT model, which predicts a fluence-dependent relaxation rate in systems with a narrow energy gap when the electron-boson recombination rate or the boson relaxation time is large. This further suggests that the pseudogap in $CsCa_2Fe_4As_4F_2$ is closely linked to the dynamics of quasiparticles and their interaction with bosonic modes. To quantify the pseudogap energy scale, we followed the methodology of Liu $et\,al.$ [46], who analyzed the hidden-order gap in $URu_2Si_2$. By fitting the temperature dependence of $\tau_2$ with Eq. (3) [Fig. 2(d)], we estimated a pseudogap size of $\Delta_{PG} = 3.3 \pm 0.3$ meV for $CsCa_2Fe_4As_4F_2$. In addition, a pseudogap with a similar size was obtained by fitting the temperature dependence of the density of thermally excited quasiparticles $n_T$ with Eq. (4) above $T_c$ [inset of Fig. 2(c)]. In addition to the dominant role of the gradual opening of the pseudogap on the Fermi surface, the continuous increase of $\tau_2$ at low temperatures below $T_c$ may also be influenced by the suppression of normal state signals due to the opening of the superconducting gap and the intricate interplay of multiple relaxation processes in the multi-band system.

The question then becomes, what is the origin of this pseudogap? We believe it stems from the multi-band nature of this material, where each band has its own unique electronic properties. This allows for the intriguing possibility of band-selective superconducting fluctuations, where some bands might show stronger fluctuations than others. Several observations support this hypothesis. Transport and NMR measurements [28] have revealed anomalies below $\sim 90$ K, potentially attributable to superconducting fluctuations. Additionally, optical studies [22] have indicated band-selective clean-limit and dirty-limit superconductivity in this material. ARPES measurements [31] have confirmed the presence of multiple electron and hole pockets, consistent with a multi-band system. Based on this evidence, we suggest that the observed pseudogap originates predominantly from superconducting fluctua-tions in a specific band, likely the clean-limit band. This implies that the strength of superconducting fluctuations varies among bands. Similar behavior has been reported in other multi-band superconductors, such as Ba(Fe$_{1-x}$Co$_x$)$_2$As$_2$ [47], further supporting our proposition. While other explanations for the pseudogap exist, such as competing orders [2], the absence of any magnetic or charge order in CsCa$_2$Fe$_4$As$_4$F$_2$ makes these alternatives less likely. It is important to note that in a multi-band system, the observed relaxation time may be a combination of several relaxation processes [48]. This can lead to a non-trivial temperature dependence of the relaxation time, potentially masking or mimicking features associated with the pseudogap phase.

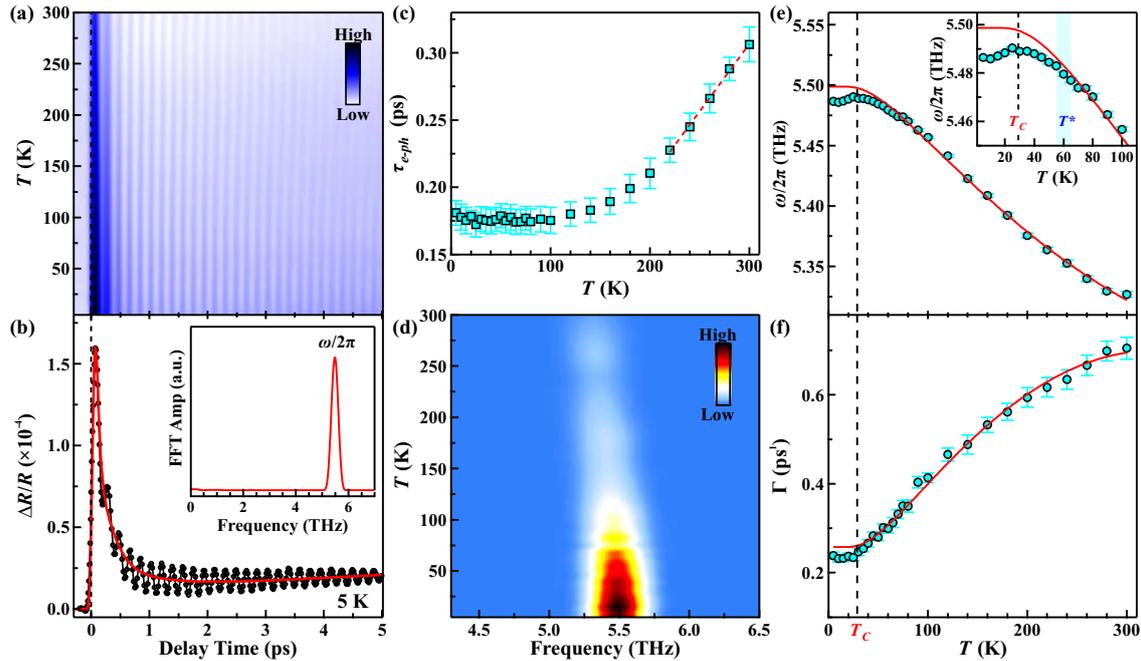

FIG. 4. (a) 2D color map of $\Delta R/R$ as a function of temperature and delay time under a pump fluence of $\sim 99~\mu J/cm^2$. (b) Representative $\Delta R/R$ at 4 K, showing the decomposition into quasiparticle relaxation (red solid line) and coherent phonon oscillations. Inset: FFT spectrum revealing the $A_{1g}$ phonon mode at 5.49 THz. (c) Temperature dependence of the $e$-ph scattering time $\tau_{e-ph}$. The red dashed line represents a fit using the extended multi-temperature model (EMTM). (d) 2D map of the FFT spectrum as a function of temperature and frequency. (e) and (f) Temperature dependence of the frequency ($\omega$) and damping rate ($\Gamma$) of the $A_{1g}$ mode, respectively, extracted from fitting the oscillations. The red solid lines represent fits using the anharmonic phonon model.

Next, the role of $e$-ph coupling in superconductivity was examined. Figure 4(a) shows the temperature and time dependence of $\Delta R/R$ under a higher pump fluence ($\sim 99~\mu J/cm^2$). In addition to the quasiparticle dynamics, pronounced periodic oscillations persisting up to room temperature are observed. These oscillations, extracted by subtracting the exponential decay background, are shown in Fig. 4(b). A fast Fourier transform (FFT) analysis of these oscillations reveals a single coherent phonon mode at 5.49 THz (i.e., 22.7 meV or 183.1 cm$^{-1}$) [inset of Fig. 4(b)], consistent with the $A_{1g}$ Raman-active phonon mode associated with $c$-axis polarized vibrations of the FeAs layers [32, 49, 50]. It is worth noting that without additional theoretical calculations, such as density functional perturbation theory (DFPT), we cannot definitively determine the irreducible representation for this material, nor can we rule out the possibility of other in-plane modes contributing to the observed signal, especially given the potential for contributions from closely spaced or overlapping modes. Future studies, including the aforementioned DFPT calculations of the phonon modes and polarization-dependent Raman spectroscopy measurements, are planned and will help to address this issue more comprehensively and definitively.

The $e$-ph coupling lifetime $\tau_{e-ph}$ ($\tau_2$) exhibits a distinct temperature dependence [Fig. 4(c)], remaining nearly constant at low temperatures and increasing above $\sim 100$ K. Above high temperatures, $\tau_{e-ph}$ is proportional to the sample temperature, which is well-described by the extended multi-temperature model (EMTM) [51], commonly used for cuprates and FeSCs [52–54]. The relationship between the second moment of the Eliashberg spectral function $\lambda \langle \omega^2 \rangle$ and $\tau_{e-ph}$ can be expressed as: $\lambda \langle \omega^2 \rangle = 2\pi k_B T_l/3\hbar\tau_{e-ph}$, where $\omega$ is the phonon frequency, and $T_l$ is the lattice temperature. Usually, $T_l$ is close to the ambient temperature due to the high heat capacity of the lattice. The value of $\lambda \langle \omega^2 \rangle$ was estimated to be $\approx 2.68 \times 10^{26}$ Hz$^2$ (i.e., $\sim 52$ meV$^2$). In the absence of detailed phonon density of states information for CsCa$_2$Fe$_4$As$_4$F$_2$, a nominal $e$-ph coupling constant, $\lambda_{A_{1g}} \approx 0.23$, was estimated based on the $A_{1g}$ mode frequency. This value, along with its $T_c$, aligns with the previously reported positive correlation between $\lambda_{A_{1g}}$ and $T_c$ in other FeSCs [14].

The $A_{1g}$ phonon mode exhibits a striking temperature dependence, with a significant blueshift and intensity increase as temperature decreases [Fig. 4(d)]. To quantitatively analyze this behavior, the frequency ($\omega$) and damping rate ($\Gamma$) were extracted by fitting the damped oscillations (see Sup-plemental Material [35]). While the damping rate (Γ) of the oscillations follows the expected anharmonic behavior [55, 56], the frequency $\omega$ deviates, displaying a subtle yet crucial downturn below $T_c$ [Figs. 4(e) and 4(f)]. This softening of the $A_{1g}$ phonon below $T_c$, though small, suggests its participation in the superconducting condensate, mirroring observations in other iron-based superconductors such as FeSe and $Fe_{1.05}Se_{0.2}Te_{0.8}$ [57]. Within our experimental resolution, we also observed that $A_{1g}$ phonon frequency deviates slightly from the anharmonic prediction around the pseudogap temperature $T^*$ [inset of Fig. 4(e)]. This close connection between the phonon behavior and the pseudogap phase strongly suggests that the $A_{1g}$ phonon contributes not only to superconducting condensation but also to the formation of the pseudogap. This connection between phonon frequency anomaly and pseudogap formation was also reported in cuprate superconductors [58], further strengthening this hypothesis. These findings collectively underscore the noteworthy role of phonons in $CsCa_2Fe_4As_4F_2$. The $A_{1g}$ phonon, through its strong coupling to electrons, appears to contribute to both the superconducting condensate and the pseudogap formation. This suggests that $e$-ph coupling is a factor to consider when investigating the intricate interplay between superconductivity and the pseudogap phase in this material.

In summary, ultrafast optical spectroscopy was employed to investigate the quasiparticle dynamics and coherent phonon oscillations in $CsCa_2Fe_4As_4F_2$ single crystals. The temperature dependence of quasiparticle relaxation revealed the presence of a pseudogap with a magnitude of $\Delta_{PG} = 3.3 \pm 0.3$ meV below $T^* \approx 60$ K, which precedes the emergence of a superconducting gap of $\Delta(0) = 6.6 \pm 0.4$ meV below $T_c$. Notably, a coherent $A_{1g}$ phonon mode exhibited an anomalous frequency downturn below $T_c$, and the frequency variations deviated from anharmonicity precisely at the pseudogap temperature ($T^*$). These observations suggest a coupling between this mode and both the superconducting condensate and the pseudogap formation. The electron-phonon coupling constant for this $A_{1g}$ mode was estimated to be $\lambda_{A_{1g}} \approx 0.23$. These results offer valuable insights into the interplay of the pseudogap, electron-phonon interactions, and the superconducting pairing mechanism in iron-based superconductors.

This work was supported by the National Natural Science Foundation of China (Grants No. 92265101 and No. 12074436), the National Key Research and Development Program of China (Grants No. 2022YFA1604204), and the Science and Technology Innovation Program of Hunan Province (2022RC3068).

* Corresponding author: jqmeng@csu.edu.cn